# AI Literacy in UAE Libraries: Assessing Competencies, Training Needs, and Ethical Considerations for the Digital Age

Zafar Imam Khan, Learning Resources Manager, Hamdan Bin Mohammed Smart University, Dubai, United Arab Emirates, Email: zafarimamkhan@gmail.com, https://orcid.org/0000-0003-2081-0951

## Abstract

The study explores the current state of artificial intelligence (AI) literacy levels among library professionals employing a quantitative approach consisting of 92 surveys of LIS professionals in the United Arab Emirates (UAE). Findings of the study revealed the presence of strong cognitive competencies, while there were gaps observed in behavioral and normative competencies, especially related to AI biases, AI-powered learning, and ethical considerations. There was a disconnect observed between the perceived importance of AI skills and the effectiveness of the current training programs.

## Introduction

Generative AI has created massive disruption in all sectors, such as manufacturing, services, agriculture, medicine, and education, and has transformed a range of operations and services. Libraries are transforming and gearing up to harness the power of AI, which can enhance efficiency, accessibility, and personalization of services; thereby reshaping the traditional library landscape. This transformation has been observed in several of the traditional library services as AI is automating routine tasks such as cataloguing and classification of collections, and enhancing search functionalities and information retrieval, thereby creating a much more accurate and organized library system while librarians have more time to focus on intellectually stimulating activities (Preethi, 2024). There is a race to integrate AI into library services at a global level, and this has presented both opportunities and challenges in terms of AI literacy among library professionals. AI literacy involves understanding of AI tools, their applications, and ethical considerations surrounding their use. AI in libraries has grown into prominence but has left a significant gap in terms of AI literacy, which needs to be addressed for effective utilization of AI technologies and to educate library users. This study intends to explore the existing status of the AI literacy level of professionals working in different libraries in the United Arab Emirates (UAE). There are no studies available that focus on the UAE or the Gulf Cooperation Council (GCC) region, although very few studies related to generative AI and libraries have focused on the developed world. Most librarians lack the necessary skills and have very limited or no knowledge or understanding of basic AI concepts, algorithms, and ethical issues concerning AI technologies (Kizhakkethil & Perryman, 2024). The librarian's role has been evolving over the years due to technological advancement, and now AI literacy is a crucial skill that can bridge the gap between digital literacy and ethical AI usage (Diyaolu et al., 2024). There is a growing concern related to the impact of AI on the media landscape and the lack of AI skills among library professionals (Andersdotter, 2023). Chigwada (2024) advocated for developing frameworks for AI digital literacy courses for LIS professionals in academic libraries to address these gaps, whereas Andersdotter (2023) recommended AI-related courses for librarians, as it has shown promising results. It is also important that while proposing a framework for AI literacy courses in academic libraries, there needs to be a collaboration among library professionals, students, faculty, and Information communication technology (ICT) staff, as it will address constraints and promote safe AI practices (Chigwada, 2024; Kizhakkethil & Perryman, 2024).

**Research Problem**

There is not much literature available related to AI literacy in the Middle East in general, as compared to the other regions of the world. This study is intended to create a roadmap and a framework for AI literacy, specifically in the context of academic libraries in the UAE and can also be applied in other types of libraries throughout the Middle Eastern regions. Several studies are based in Western, South Asian, or African contexts, which cover various aspects and elements of AI literacy, competencies, and ethical consideration in libraries (Ali & Richardson, 2025; Hossain et al., 2025; Kautonen & Gasparini, 2024; Lo, 2024; Monyela & Tella, 2024). Current literature focusses on the need for AI literacy among Library & Information Science (LIS) professionals but there is insufficient empirical data related to actual competency level, preparedness level of library staff and specific training needs in the context of UAE to integrate AI into various library services and operations (Ali & Richardson, 2025; Kautonen, & Gasparini, 2024; Lo, 2024). There is also a lack of research related to unexplored ethical considerations (Mannheimer et al, 2024; Monyela & Tella, 2024), and no tailored frameworks and guidelines exist in the context of the UAE. Most of the framework and guidelines for AI literacy are generic or are adapted from other regions with little evidence of adaptation or development for academic libraries in the UAE due to a lack of studies to investigate (Chee et al, 2024; Lo, 2023). UAE public libraries are exploring and updating the current information literacy programs to develop digital awareness, especially related to AI, and focusing on the need for responsible and safe use of AI tools and technologies (Abd Al Samad et al, 2024). Najdawi (2020) emphasizes the need to consider AI adoption strategies across different organizations from both public and private sectors, involving different domains in the UAE, but faces challenges in implementation due to the lack of a framework for AI integration. This gap impedes the ability of libraries to fully adopt, utilize, and responsibly integrate AI tools and technologies.

## Objectives of Research

1. To measure the current level of AI literacy among library and information professionals, focusing on cognitive, behavioral, and normative competencies.
2. To evaluate the effectiveness of existing AI literacy training programs for library professionals.
3. To examine the ethical considerations and challenges associated with AI integration in library services.

## Literature Review

### AI Literacy and Key Competencies

Generative AI creates personalized user experience as machine learning algorithms can analyze the user information seeking behavior and based on that provide personalized recommendations, which in turn enhance user engagement and satisfaction (Amalia et al, 2024). AI in libraries can improve service delivery through advanced data analytics by understanding and responding to user needs in real time (Okwu et al, 2024). One of the key applications of AI technologies is providing AI-powered assistive technologies such as translation services, text-to-speech, and navigation systems, which can significantly enhance accessibility to people of determination and disabilities. Virtual assistants such as chatbots utilizing natural language processing are assisting patrons in navigating library systems and answering queries (Chauhan, 2024). Promoting AI literacy will be integral to the professional development of library professionals as they will be taking on the task of ensuring ethical use of AI, data privacy, and aligning AI tools and applications with library values of inclusivity and accessibility (Okwu et al, 2024; Tzanova, 2024). Lo (2024) emphasized hands-on training and reskilling programs, while Scotti & Beltran (2024) advocated for workshops like "Python for all" for library professionals. Deshen & Aharony (2024) suggested promoting AI acceptance and usage among the library professionals, while Sen (2024) advised integration of AI tools like ChatGPT and other library services. Various other studies, such as Garnier et al (2024), point towards addressing ethical and societal implications and concerns to develop AI literacy among library professionals. Kautonnen & Gasparini (2024) advocated for the B-wheel model, which is related to behavioral changes and is inspired by design thinking and has the capabilities to build AI competencies among library and information professionals in academic libraries. Alam et al (2024) pointed out a few barriers and challenges of implementation of these strategies, which include resistance to change, budgetary constraints, and the need for enhanced expertise to ensure successful integration of AI literacy programs in library settings. AI literacy can be defined as the ability to understand, engage, and critically evaluate AI-related tools and technologies. It encompasses a wide range of competencies to navigate and interact with various AI tools and systems effectively and responsibly (Almatrafi et al, 2024; Yuan et al, 2024; Zhang et al., 2024). The key components of AI literacy are cognitive competencies, behavioral competencies, and normative competencies. The cognitive competencies include understanding of AI features and processing and algorithm influences (Almatrafi et al, 2024; Yuan et al, 2024). The behavioral competencies correspond to user efficacy (Yuan et al, 2024) and application and creation (Almatrafi et al, 2024). The normative competencies comprise of ethical considerations and threat appraisal influences (Almatrafi et al, 2024; Yuan et al, 2024).

## AI Literacy, Critical Thinking, and Lifelong Learning

Hollands & Breazeal (2024) assessed several STEAM projects in various schools in Spain and found that evaluation and critical thinking skills are the key components of AI literacy. They determined that students and teachers involved in STEAM projects should be well-equipped with STEM-based competencies and instructional design knowledge to critically assess AI tools and technologies while identifying their key strengths and limitations. AI literacy also involves critical thinking skills as it assists people in identifying the challenges posed by AI technologies and finding innovative solutions within their work environment (Allen & Kendeou, 2023). The nature of AI is always evolving at a rapid pace, and to cope with these technologies, the curriculum from different disciplines needs to emphasize continuous learning and adaptability, and promote a culture of lifelong learning. Information professionals need to always keep on learning and stay updated with the latest developments in AI and best practices (Breazeal et al, 2023; Kizhakkethil & Perryman, 2024). Information professionals need to enhance and continue their professional development journey by undergoing new AI-related courses, training, workshops, and participation in various AI-related projects (Kautonen & Gasparini, 2024). Another way of continuous learning involves participating in learning circles where information professionals can discuss and collaborate on AI-related content and its application and integration in library services. This will foster a learning environment and build confidence among library professionals (Andersdotter, 2023). UNESCO's AI competency framework for teachers can guide the development of AI literacy through structured pathways for learning and assessment, as it can help in differentiating between use and misuse in education and develop knowledge, skills, and values for responsible use of AI technologies in classrooms (Faruqe et al, 2022; Mutawa & Shruthi, 2025). Evaluation and feedback mechanisms need to be incorporated to measure the effectiveness of AI technologies and information sources to ensure that learners are acquiring the necessary AI skills and competencies (Allen & Kendeou, 2023; Holland & Breazeal, 2024). Having hands-on experience with these tools will equip the learners to use them for creating bibliographic records and enhancing library services (Snow et al, 2024). Machine learning technologies and data analytics are used for important tasks such as classification of resources, clustering, and visualization of data. Acquiring these skills will provide a competitive advantage in managing big data and improving the overall library operations and services (Luca et al, 2022).

## Key Challenges, Barriers, and Concerns

Information professionals must understand and get familiar with AI concepts such as generative AI, machine learning, and natural language processing to make informed decisions about AI tool integration with various library services and operations (Pickett & Pennington, 2024). Library and information professionals are exploring various AI-powered tools to upscale and enhance their services in information retrieval, personalized and adaptive learning experiences, and tailored recommendations (Bhuvaneswari & Rajakumar, 2024; Molopa et al, 2024). The ability to convey complex technical information in an accessible manner will create rapid awareness related to AI literacy among users (Zhang et al., 2024). Collaborative skills among information professionals require them to work in tandem with AI and humans, and this machine-human interaction can foster teamwork, especially in AI-driven projects (Du, 2024). One of the key challenges associated with AI is ethics, and it is of paramount importance that information professionals should have a higher understanding of the ethical implications of AI and address these challenges, which are not limited to data privacy, hallucinations, biasedness, and unreliable and misinformation (Bhowmick et al, 2024). Information professionals need to work on their personalized learning and digital literacy with a growth mindset, as AI tools can facilitate personalized and adaptive learning (Parra-Valencia & Massey, 2023). Information professionals need to collaborate with educators to transition from just LIS professional development to developing an AI literacy module for students that is interdisciplinary in approach and ensures that students and faculty members also have the necessary skills and knowledge to thrive in an AI-driven environment (Merceron & Best, 2024).

Libraries are facing challenges in terms of budget constraints in investing in information technology infrastructure to support AI applications and their integration into library services (Chatikobo & Pasipamire, 2024). In an educational environment, especially in higher education, there needs to be interdisciplinary collaboration between the library professionals, educators, and other stakeholders for developing a comprehensive AI literacy program (Merceron & Best, 2024). Lo (2024) identified that academic library professionals in the United States have a modest self-rated understanding of AI concepts, have very little hands-on experience with AI tools, and have notable gaps in knowledge related to ethical implications and collaborative AI practices. Human-AI collaboration is critical to the success of AI implementation and integration, as information professionals need to collaborate with various AI systems and understand their limitations and strengths to enhance human capabilities. It is also important to understand that human-centric AI values are maintained to ensure AI tools and applications serve to augment human decision-making rather than to replace it (Yanyi, 2024). Integration of AI information literacy (IL) into all existing IL programs will enhance students' and users' capabilities in searching, retrieving, evaluating, and using information effectively in an AI-driven environment (Carroll & Borycz, 2024; Ndungu, 2024). Information literacy delivery professionals need to upskill themselves to teach AI literacy to users, making them understand and navigate AI tools and technologies (Andersdotter, 2023). Data literacy is a key skill that information professionals need to acquire to understand data management and big data technology awareness. This will enable them to manage and analyze large datasets, understanding data security and ensuring the integrity of data used in AI (Jinghua, 2021; Daxing, 2021). Knowledge of AI algorithm literacy and algorithm transparency among information professionals is important as it enables them to know how AI algorithms work, identify potential biases, and interpret their outputs. Furthermore, advocating for algorithm transparency in AI algorithms ensures that they are clear and accountable (Ndungu, 2024). Another key skill professionals need to acquire is digital literacy, which will assist them in navigating misinformation, such as deep fakes, which are produced by AI tools, and will ensure the promotion of responsible digital citizenship and ethical use of AI technologies (Adetayo, 2021).

## AI Literacy Frameworks for Academic Libraries

Many frameworks exist on AI literacy in general and in different contexts, but two studies have proposed an AI literacy framework for LIS professionals that particularly stand out as they focus on academic libraries, whereas most AI literacy frameworks are related to education in general. Lo (2025) proposed a framework that comprises five core components and three cross-cutting themes that permeate the framework. The core components include technical knowledge, ethical awareness, critical thinking, practical skills, and societal impact, while cross-cutting themes include human-AI collaboration, lifelong learning and adaptation, and equity and access. This study also advocated a dual implementation approach involving professional development and training for staff and community engagement through collaboration with educators across disciplines and integration with the current information literacy program. Hossain et al. (2025) proposed a human-centered AI Literacy framework (Figure 1 below), which conceptualized AI literacy through four interconnected dimensions, which include Foundational AI literacy, Operational AI literacy, Transformational AI literacy, and Critical AI literacy, revolving around AI literacy skill classified along normative and cognitive axes.

**Figure 1: Human-Centered AI Literacy Framework**

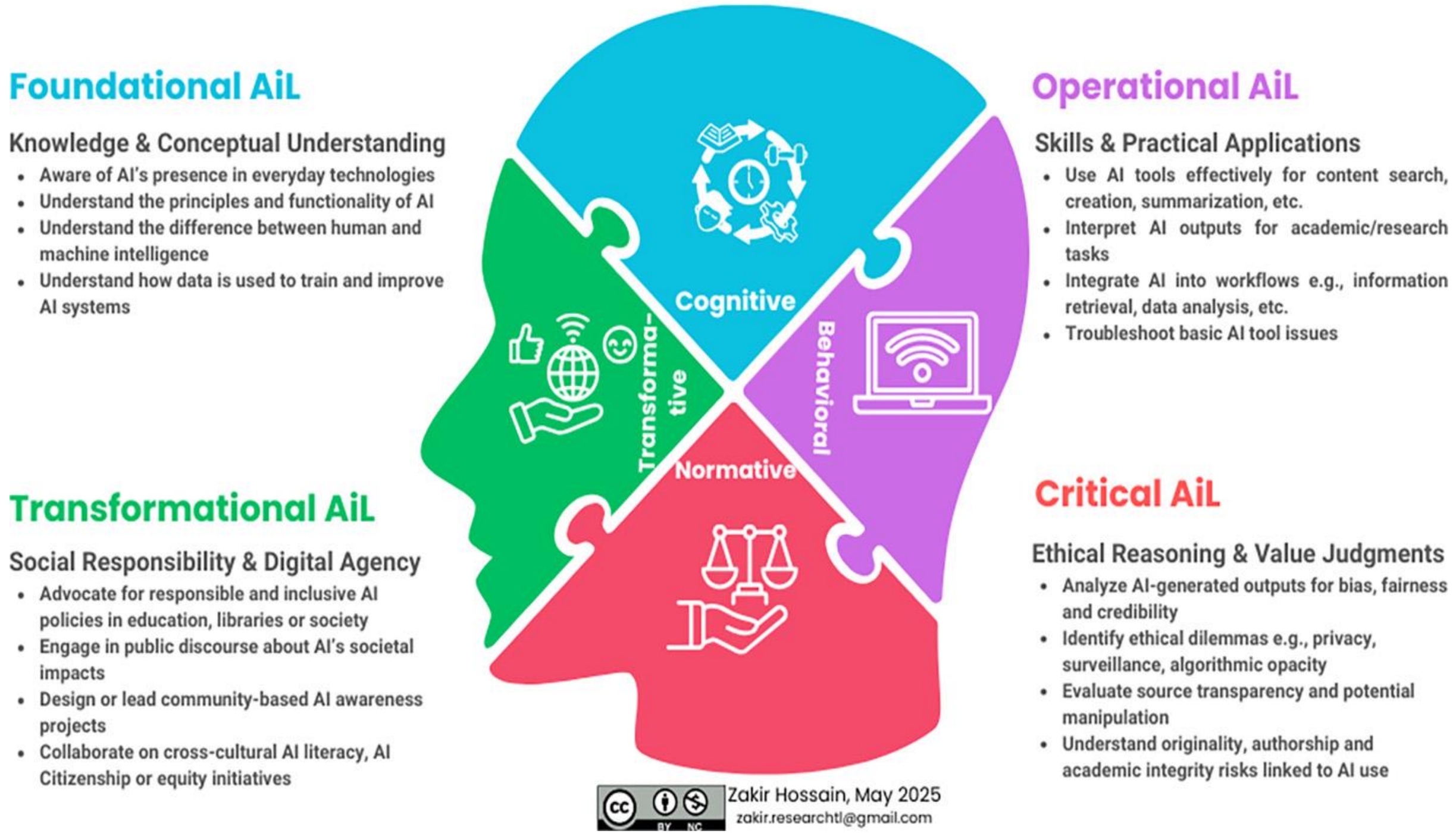

**Source: (Hossain et al., 2025)**

Figure 2 below shows the proposed framework on AI literacy for Library & Information Science professionals based on the review of the existing literature. The conceptual framework below outlines the key concepts, variables, and their relationships that guide the design of this study.

**Figure 2: Proposed Conceptual Framework on AI Literacy for LIS professionals.**

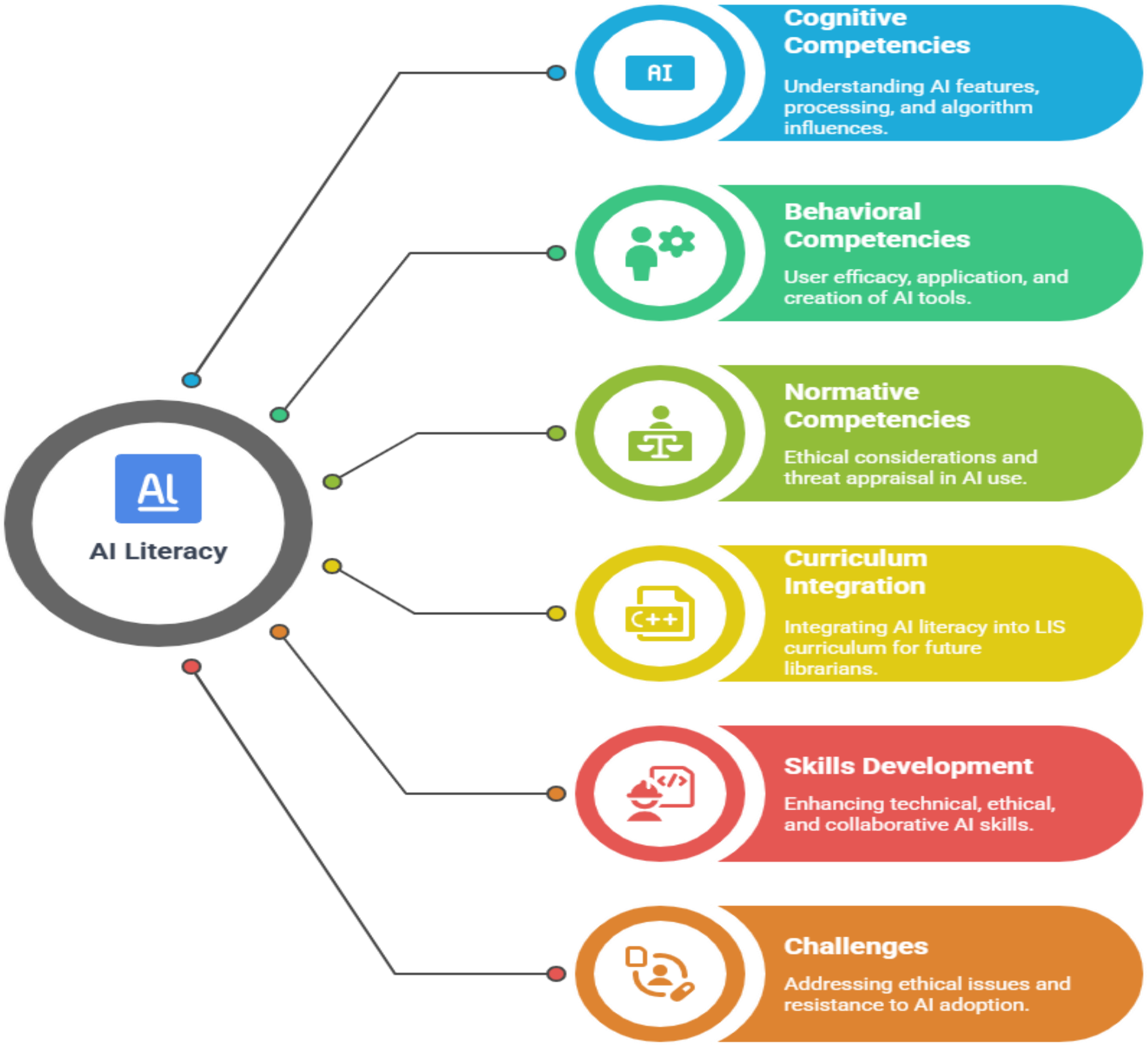

Source: Author self-generated

**Research Methodology**

This study involves quantitative methodology employing a survey to assess the need to build AI literacy among library professionals in the UAE. The study primarily targeted a diverse sample population of library professionals from universities, colleges, schools, and public libraries across the country. The survey instrument was developed and designed to measure various levels of AI competencies in terms of behavioral, cognitive, and normative competencies, AI literacy skills and knowledge, effectiveness of AI training, and ethical considerations and challenges associated with AI. The survey instrument was developed using the literature covered, and each of these variables was identified by critically evaluating the literature review, and was not focused on any specific study or studies. The survey consists of five demographic data-related questions with multiple choice options and six constructs, which include 30 variables in the form of Likert Scale items, measuring and evaluating participants' attitudes, skills, and knowledge towards AI application in library services. The constructs and variables were developed and identified through the literature review and critically examined in the context of AI literacy. The structured online survey was designed using Microsoft Forms, and the response collection link was distributed following random sampling, and further snowball sampling techniques were used for a broader reach. Participants were requested to share the survey link within their professional networks, which helped in expediting the data collection process. The study adhered to a rigorous ethical framework, ensuring compliance with all established guidelines and protocols under the ethical standards of the Declaration of Helsinki. Participants in the study were provided with informed consent forms, clearly outlining the purpose of the study and measures taken to ensure confidentiality during the research process. All the data collected were managed securely, and respondents' anonymity was strictly maintained throughout the study. A total of 92 responses were collected from library & information science professionals across the UAE. The statistical analysis included descriptive statistical analysis to summarize demographic information and pattern of response, correlation analysis to examine the relationships between variables, independent sample T-test to compare variables between genders.

**Data Analysis and Findings**

The data analysis involved employing a range of statistical techniques to examine the dataset comprehensively.

Table 1 below shows the demographic data distributions of the independent variables. The gender distribution among the respondents shows higher proportions of male librarians (60.87%) as compared to female librarians (39.13%). The data related to age groups suggest that most of the respondents are between 35-54 and comprise almost 80% of the total respondents, while there was no respondent found in the 18-24 age group. The age group, 55 and above, is relatively small (13.04%) but suggests that a significant number of older adults participated in this study. The average age of the respondents is 42.5 years. The largest group (57.61%) of library professionals are working in higher education, comprising university and college libraries, while school libraries account for 32.61% and special libraries 9.78% of the respondents. There was no representation of public library professionals in this study, which can be one of the limitations of this study. As for the data related to the highest level of education, it shows that the majority of library professionals have a master's degree (55.43%), followed by PhDs (27.17%) and Bachelors (17.39%) of the respondents. It is also observed that the majority of library professionals (60.87%) have more than 15 years of experience, with an average of 12.8 years of experience, which suggests a very experienced sample size of population in this study, which can contribute towards a more in-depth understanding of the importance and need of AI literacy among library professionals.

**Table 1: Demographic Details of the Respondents**

| Demographic Variable | Category | Frequency | Percentage (%) |
|---|---|---|---|
| **Gender** | **Male** | 56 | 60.87% |
| | **Female** | 36 | 39.13% |
| **Age** | **18-24** | 0 | 0% |
| | **25-34** | 10 | 10.87% |
| | **35-44** | 36 | 39.13% |
| | **45 -54** | 34 | 36.96% |
| | **55 and above** | 12 | 13.04% |
| **Type of Library** | **Higher Education (Colleges/Universities)** | 53 | 57.61% |
| | **School Library** | 30 | 32.61% |
| | **Special Library** | 9 | 9.78% |
| | **Public Library** | 0 | 0% |
| **Highest Level of Education** | **Bachelor's** | 16 | 17.40% |
| | **Master's** | 51 | 55.43% |
| | **PhDs** | 25 | 27.17% |
| **Years of Experience** | **0 to 5** | 6 | 6.52% |
| | **6 to 10** | 20 | 21.74% |
| | **11 to 15** | 10 | 10.87% |
| | **16 and above** | 56 | 60.87% |

Table 2 below clearly shows that most librarians surveyed depict a positive self-assessment of their AI literacy and understanding and have responded with agree or strongly agree with the statement across all variables in this construct. Almost 85.87% (Agree or Strongly Agree) of the total respondents can explain basic principles of AI (M=4.09, SD= 0.78) to others. This suggests that most of the librarians are confident in their ability to communicate fundamental AI concepts with others. The understanding of AI technologies (M=3.80, SD=0.81) among librarians still suggests a positive inclination, but the lower mean value as compared to other variables might indicate that they feel less confident when it comes to in-depth understanding of specific AI technologies as compared to general principles. Familiarity with AI tools (M=4.16, SD=0.71) data suggests that the majority (85.87%) of librarians are very confident in their knowledge of popular AI tools used in the library. As for AI applications in libraries (M=4.00, SD=0.72), there is a positive sign that 77.26% of respondents either strongly agree or agree, suggesting that librarians are equipped to identify potential AI applications in various library services and how they can be integrated into their operations. The librarians (79.35%) also seem to have a high awareness level and understanding of the limitations of AI (M=3.91, SD=0.80).

**Table 2: Current Level of AI Literacy (Cognitive Competencies)**

| Statement | Strongly Disagree (N) | Disagree (N) | Neutral (N) | Agree (N) | Strongly Agree (N) | Total (N) / % | Mean (M) | Standard Deviation (SD) |
|---|---|---|---|---|---|---|---|---|
| Explain the basic principles of artificial intelligence to others. | 1 (1.09%) | 3 (3.26%) | 9 (9.78%) | 53 (57.61%) | 26 (28.26%) | 92 (100%) | 4.09 | 0.78 |
| Understand the differences between various AI technologies (e.g., machine learning, natural language processing, generative AI) | 1 (1.09%) | 4 (4.35%) | 23 (25.00%) | 48 (52.17%) | 16 (17.39%) | 92(100%) | 3.80 | 0.81 |
| Familiar with popular AI tools used in library services (e.g., ChatGPT, Gemini etc) | 0 (0%) | 2 (2.17%) | 11 (11.96%) | 49 (53.26%) | 30 (32.61%) | 92 (100%) | 4.16 | 0.71 |
| Identify potential applications of AI in library and information services. | 0 (0%) | 2 (2.17%) | 18 (19.57%) | 50 (54.35%) | 22 (22.91%) | 92 (100%) | 4.00 | 0.72 |
| Understand the limitations of current AI technologies. | 1 (1.09%) | 5 (5.43%) | 13 (14.13%) | 55 (59.78%) | 18 (19.57%) | 92 (100%) | 3.91 | 0.80 |

Table 3 below shows that overall, there is a positive sentiment, but the responses are distributed across the agreement scale, which indicates a wider range of engagement among librarians with AI. Using AI for creating bibliographic records (M=3.29, SD=1.05) indicates that there is a moderate experience, and there is significant variation. A little more than half of the respondents (53.27%) agree or strongly agree with this statement. As for the AI-powered data analytics (M=3.46, SD=0.94) variable, the trend is the same, and just a little more than half of the respondents (56.52%) have this competency. When it comes to the integration of AI into library services (M=3.72, SD=0.84), respondents showed an upward swing as almost 69.56% of respondents either agree or strongly agree. This suggests that the majority of librarians are comfortable when it comes to integrating AI tools into existing library services. Using AI for information retrieval (M=3.80, SD=0.78), also showed a high level of confidence among the respondents, with almost 78.26% of respondents agreeing or strongly agreeing. This suggests that most librarians show a high degree of confidence when it comes to using AI to enhance information retrieval processes, and this was the highest-ranking variable in this construct. As for AI-powered personalized learning (M=3.23, SD=1.08), a decline in confidence level is observed, and almost less than half (48.92%) agree or strongly agree on this competency level.

**Table 3: Current Level of AI Literacy (Behavioral Competencies)**

| Statement | Strongly Disagree (N) | Disagree (N) | Neutral (N) | Agree (N) | Strongly Agree (N) | Total (N) / % | Mean (M) | Standard Deviation (SD) |
|---|---|---|---|---|---|---|---|---|
| Experience using AI tools for creating bibliographic records. | 5 (5.43%) | 19 (20.65%) | 19 (20.65%) | 42 (45.65%) | 7 (7.62%) | 92 (100%) | 3.29 | 1.05 |
| Use AI-powered data analytics tools to improve library operations. | 1 (1.09%) | 17 (18.48%) | 22 (23.91%) | 43 (46.74%) | 9 (9.78%) | 92 (100%) | 3.46 | 0.94 |
| Comfortably integrating AI tools into existing library services. | 1 (1.09%) | 8 (8.7%) | 19 (20.65%) | 52 (56.52%) | 12 (13.04%) | 92 (100%) | 3.72 | 0.84 |
| Use AI to enhance information retrieval processes in my library. | 1 (1.09%) | 7 (11.96%) | 12 (13.04%) | 61 (66.3%) | 11 (11.96%) | 92 (100%) | 3.80 | 0.78 |
| Designed AI-powered personalized learning experiences for library users. | 6 (6.52%) | 20 (21.74%) | 21 (22.82%) | 37 (40.22%) | 8 (8.7%) | 92 (100%) | 3.23 | 1.08 |

Table 4 below shows the current level of AI literacy among librarians in terms of normative competencies encompassing critical evaluation and ethical implications of AI in the context of a library. Identifying potential AI biases is a key challenge, and the data shows that 55.44% of the respondents agree or strongly agree that they can identify potential biases in AI algorithms and their outputs (M=3.39, SD=0.96). This suggests a moderate confidence level in recognizing bias and that there is room for improvement. Privacy awareness (M=3.62, SD=0.87) was much better placed as 63.05% of the respondents agree or strongly agree that they are aware of the privacy implications while using AI in library services. As for evaluation of AI-generated information (M=3.60, SD=0.97), it was observed that 60.87% of respondents agree or strongly agree on their ability to critically evaluate the reliability and accuracy of AI-generated information. This suggests that librarians are well capable of assessing AI outputs critically and identifying false or “hallucinating” information. The variable related to ethical consideration (M=3.88, SD=0.83) scored the highest, with 76.09% respondents agreeing or strongly agreeing with it. This is a positive sign, and it implies that librarians are highly aware of the ethical aspects of using AI in information retrieval and dissemination. When it comes to the question of AI accountability (M=3.78, SD=0.83), it was also rated highly, with 72.83% respondents agreeing or strongly agreeing to it. This suggests that librarians have a good understanding of the concept of AI accountability and can explain it to library users.

**Table 4: Current Level of AI Literacy (Normative Competencies)**

| Statement | Strongly Disagree (N) | Disagree (N) | Neutral (N) | Agree (N) | Strongly Agree (N) | Total (N) / % | Mean (M) | Standard Deviation (SD) |
|---|---|---|---|---|---|---|---|---|
| Identify potential biases in AI algorithms and their outputs. | 2 (2.17%) | 18 (19.57%) | 21 (22.82%) | 44 (47.83%) | 7 (7.61%) | 92 (100%) | 3.39 | 0.96 |
| Aware of the privacy implications of using AI in library services. | 3 (3.26%) | 5 (5.43%) | 26 (28.26%) | 48 (52.17%) | 10 (10.88%) | 92 (100%) | 3.62 | 0.87 |
| Critically evaluate the reliability and accuracy of AI-generated information. | 3 (3.26%) | 9 (9.78%) | 24 (26.09%) | 42 (45.65%) | 14 (15.22%) | 92 (100%) | 3.60 | 0.97 |
| Understand the ethical considerations of using AI in information retrieval and dissemination. | 2 (2.17%) | 3 (3.26%) | 17 (18.48%) | 52 (56.52%) | 18 (19.57%) | 92 (100%) | 3.88 | 0.83 |
| Explain the concept of AI accountability to library users. | 3 (3.26%) | 2 (2.17%) | 20 (21.74%) | 54 (58.70%) | 13 (14.13%) | 92 (100%) | 3.78 | 0.83 |

Table 5 below relates to essential AI literacy skills and knowledge areas. This is one of the key constructs, and it deals with the variables related to the importance and need of AI literacy skills in the field. The first variable related to understanding of AI algorithms (M=3.92, SD=0.87) showed a high positive confidence, with 75% of respondents agreeing or strongly agreeing on this. This suggests that understanding of AI algorithms and machine learning principles is crucial for the library & information science (LIS) professionals, and technical AI knowledge is particularly important. As for the practical AI experience (M=4.09, SD=00.86), the data shows that 82.61% of respondents agree or strongly agree that practical experiences with AI tools need to be the core component of Library & Information Science education, and hands-on experience and learning on AI technologies is strongly recommended. The variable related to data management knowledge (M=4.21, SD=0.60) is the highest-ranked variable in the construct, with 90.22% agreeing or strongly agreeing on this. This implies that there is near unanimity among the respondents in relation to the importance of data management skills for working with AI in libraries. The critical thinking for AI evaluation (M=4.17, SD=0.73) was equally highly rated with 88.04% respondents either agreeing or strongly agreeing that critical thinking skills for evaluation of AI tools are crucial and vital for LIS professionals, as it can equip them with strong analytical skills for evaluation of AI tools and technologies. As for the societal impact of AI (M=4.02, SD=0.84), respondents emphasized its importance too as 78.26% agreed or strongly agreed on its inclusion as part of the LIS curriculum, suggesting a recognition of the wider implications of AI beyond just the technical aspects of it.

**Table 5: Essential AI Literacy Skills and Knowledge Areas**

| Statement | Strongly Disagree (N) | Disagree (N) | Neutral (N) | Agree (N) | Strongly Agree (N) | Total (N) / % | Mean (M) | Standard Deviation (SD) |
|---|---|---|---|---|---|---|---|---|
| Understanding AI algorithms and machine learning principles is crucial for LIS professionals. | 1 (1.09%) | 5 (5.43%) | 17 (18.48%) | 46 (50.00%) | 23 (25.00%) | 92 (100%) | 3.92 | 0.87 |
| Practical experience with AI tools should be a core component of LIS education. | 2 (2.67%) | 1 (1.09%) | 13 (14.13%) | 47 (51.09%) | 29 (31.52%) | 92 (100%) | 4.09 | 0.86 |
| Knowledge of data management and analysis is essential for working with AI in libraries. | 0 (0%) | 0 (0%) | 9 (9.78%) | 55 (59.78%) | 28 (30.44%) | 92 (100%) | 4.21 | 0.60 |
| Critical thinking skills for evaluating AI tools are vital for LIS professionals. | 1 (1.09%) | 1 (1.09%) | 9 (9.78%) | 51 (55.43%) | 30 (32.61%) | 92 (100%) | 4.17 | 0.73 |
| Understanding the societal impact of AI should be part of the LIS curriculum. | 1 (1.09%) | 3 (3.26%) | 16 (17.39%) | 45 (48.91%) | 27 (29.35%) | 92 (100%) | 4.02 | 0.84 |

Table 6 below shows the effectiveness of existing AI Literacy training among librarians provided by their institutions, indicating a varying level of satisfaction and exposure with AI training programs. The first variable in this construct was related to participation in AI training (M=3.27, SD=1.03), with nearly half of the respondents (48.92%) agreeing or strongly agreeing on their participation in AI literacy training programs specifically designed for library & information professionals. Regarding the comprehensive nature of the training program (M=3.23, SD=0.96), only 39.13% agreed or strongly agreed that the AI literacy training they received was comprehensive and covered all three aspects, including cognitive, behavioral, and normative competencies. This indicates a potential gap in the training content. As for the practical application of the training program (M=3.25, SD=0.94), only 41.31% of the respondents agreed or strongly agreed on the effectiveness of the use of AI tools in their job. This suggests that a moderate level of effectiveness of training in practical skill development was observed. Training addresses ethical considerations (M=3.32, SD=1.00) and followed a similar trend, with only 46.74% agreeing or strongly agreeing on this aspect. This suggests there is a need for improvement and resources required to cover the ethical aspects of AI training. The variable related to confidence in applied knowledge (M=3.38, SD=0.90) also showed results on similar lines, with 46.74% of the respondents agreeing or strongly agreeing that they feel confident while applying the knowledge gained from the AI literacy training in their professional work.

**Table 6: Effectiveness of Existing AI Literacy Training**

| **Statement** | **Strongly Disagree (N)** | **Disagree (N)** | **Neutral (N)** | **Agree (N)** | **Strongly Agree (N)** | **Total (N) / %** | **Mean (M)** | **Standard Deviation (SD)** |
|---|---|---|---|---|---|---|---|---|
| Participated in AI literacy training programs specifically | 4 (4.35%) | 20 (21.74%) | 23 (25.00%) | 37 (40.22%) | 8 (8.70%) | 92 (100%) | 3.27 | 1.03 |
| designed for library professionals. | | | | | | | | |
| The AI literacy training I received was comprehensive and covered all necessary aspects. | 3 (3.26%) | 17 (18.48%) | 36 (39.13%) | 28 (30.43%) | 8 (8.70%) | 92 (100%) | 3.23 | 0.96 |
| The training effectively prepared me to use AI tools in my daily work. | 3 (3.26%) | 16 (17.39%) | 35 (38.04%) | 31 (33.70%) | 7 (7.61%) | 92 (100%) | 3.25 | 0.94 |
| The training addressed ethical considerations of AI use in libraries. | 4 (4.35%) | 15 (16.30%) | 30 (32.61%) | 34 (36.96%) | 9 (9.78%) | 92 (100%) | 3.32 | 1.00 |
| Confidently applying the knowledge gained from AI literacy training in my professional practice. | 2 (2.17%) | 12 (13.04%) | 35 (38.04%) | 35 (38.04%) | 8 (8.70%) | 92 (100%) | 3.38 | 0.90 |

Table 7 below suggests that librarians have shown significant concerns related to several aspects and ethical implications of AI use in libraries. They have shown confidence in addressing these issues and have highlighted the need for clear ethical guidelines surrounding this. The first variable related to privacy concerns (M=3.76, SD=0.81), 66.30% respondents agreed or strongly agreed that they were concerned about potential privacy breaches while using AI in library services, which indicates a high level of awareness about privacy issues. With regards to bias amplifications (M=3.84, SD=0.83), 65.21% showed similar concerns and agreed or strongly agreed on AI potentially perpetuation or amplifying biases in information retrieval. This suggests that librarians have shown concerns related to equity and AI fairness. As for the transparency challenges (M=3.70, SD=0.80), they showed a similar trend with 64.13% respondents agreeing or strongly agreeing on ensuring transparency as a major challenge while using AI-powered tools in library services, which suggests that librarians have recognized the complexities in communicating AI processes to users. Confidence in addressing ethical issues showed 66.31% agreement in their ability to address ethical issues related to the use of AI in libraries. This suggests that there is a high confidence level among the librarians and self-assurance in addressing ethical challenges. As for the need for ethical guidelines (M=4.17, SD=0.79), most of the librarians (85.87%) were unanimous and agreed or strongly agreed on having a clear set of ethical guidelines and standards for AI use in libraries.

**Table 7: Ethical Considerations and Challenges**

| Statement | Strongly Disagree (N) | Disagree (N) | Neutral (N) | Agree (N) | Strongly Agree (N) | Total (N) / % | Mean (M) | Standard Deviation (SD) |
|---|---|---|---|---|---|---|---|---|
| Concerned about potential privacy breaches when using AI in library services. | 1 (1.09%) | 4 (4.35%) | 26 (28.26%) | 46 (50.00%) | 15 (16.30%) | 92 (100%) | 3.76 | 0.81 |
| Worry about the potential for AI to perpetuate or amplify biases in information retrieval. | 0 (0%) | 4 (4.35%) | 28 (30.43%) | 39 (42.39%) | 21 (22.82%) | 92 (100%) | 3.84 | 0.83 |
| Challenging to ensure transparency when using AI-powered tools in library services. | 1 (1.09%) | 5 (5.43%) | 27 (29.35%) | 47 (51.09%) | 12 (13.04%) | 92 (100%) | 3.70 | 0.80 |
| Confidence in my ability to address ethical issues related to AI use in libraries. | 1 (1.09%) | 4 (4.35%) | 26 (28.26%) | 50 (54.35%) | 11 (11.96%) | 92 (100%) | 3.72 | 0.77 |
| Clear guidelines for the ethical use of AI in library and information services. | 1 (1.09%) | 2 (2.17%) | 10 (10.88%) | 46 (50.00%) | 33 (35.87%) | 92 (100%) | 4.17 | 0.79 |

There were no statistically significant differences between male and female respondents (p-value was greater than 0.05 for all variables), suggesting that gender does not play any major role in most aspects of AI literacy and attitudes among library professionals.

Figure 3 below shows the proposed framework for AI literacy for library professionals based on the findings of this study.

**Figure 3: Proposed Framework for AI Literacy for Library Professionals**

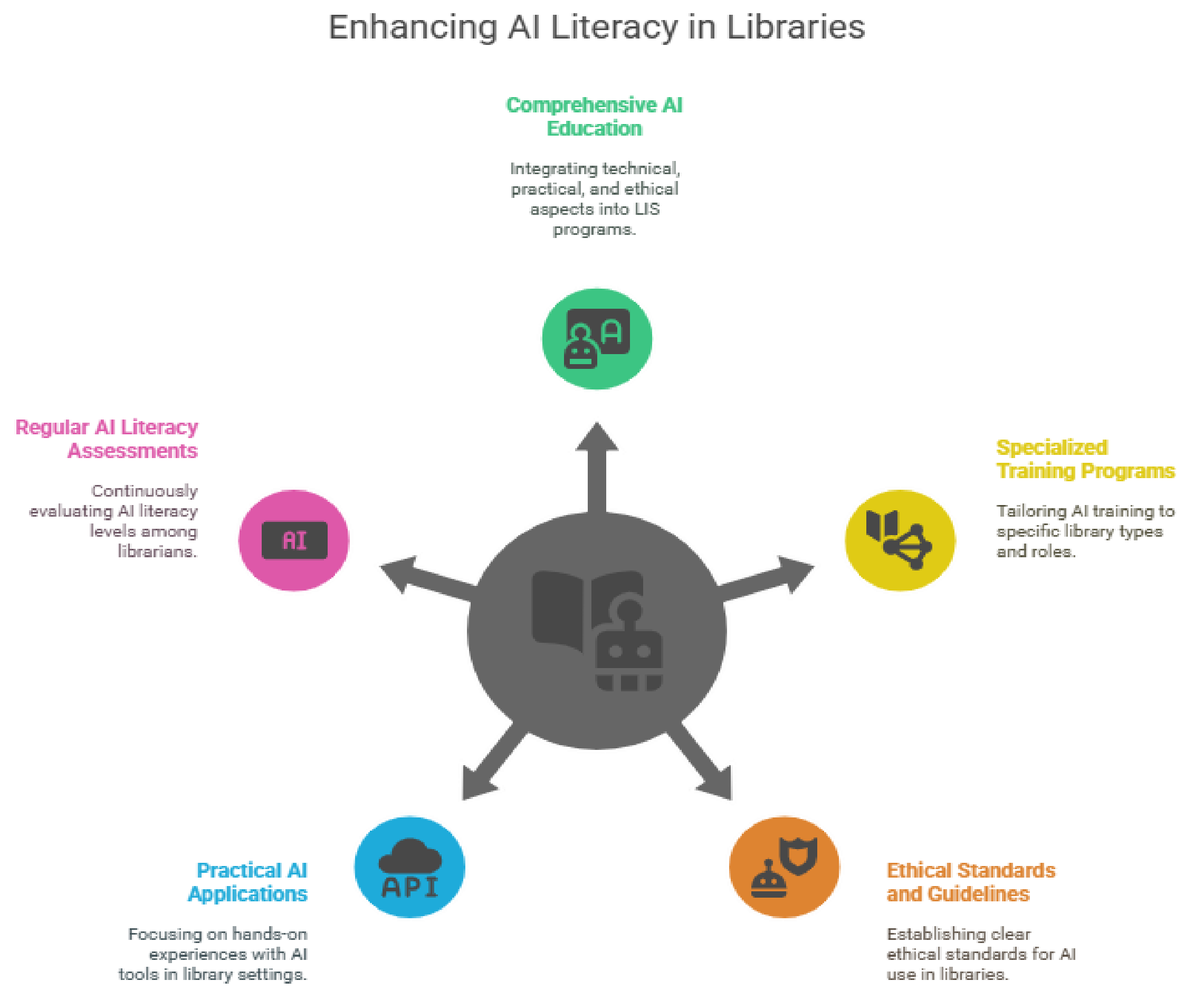

**Source: Author self-generated**

## Discussion

This study provides constructive insights into the current state of AI literacy among library & information science professionals working in different types of libraries in the UAE. The study highlights the strengths and areas of improvement and provides recommendations to address the gaps. The demographic variable highlighted the fact that the majority of the respondents were male (60.87%) and working in higher education (57.61%). The majority of the participants had an advanced level of education (55.43% Master's, 27.17% PhDs), and 60.87% had more than 15 years of experience working in libraries. The findings reveal a high level of cognitive competencies (85.87%) related to AI, where librarians demonstrated confidence in explaining basic understanding of AI principles and familiarity with AI tools (85.87%), and are in line with studies by Alam et al (2024), Lo (2024), and Mallikarjuna (2024). As for the behavioral competencies, there was moderate to high confidence observed in using AI for various library tasks, where information retrieval showed the highest level of confidence (78.26%) among the librarians. Normative competencies also showed moderate to high levels of confidence with ethical consideration (76.09%) and AI accountability (72.83%), showing the highest. However, there were notable gaps found in certain behavioral and normative competences, especially in areas related to identifying AI biases learning (Berendt, 2023; Saeidnia, 2023) and AI-powered personalized (Kaswan et al., 2024). Essential AI literacy skills constructs showed high agreement concerning variables such as the need for critical thinking for AI evaluation (88.04%) and data management knowledge (90.22%).

This research highlights the critical importance of AI literacy skills in the field of library & information science. There is unanimity and consensus among the library professionals regarding the need for comprehensive AI education in the field of LIS, which should encompass technical knowledge and understanding, practical experience, critical thinking, and ethical considerations (Chigwada, 2024; Kizhakkethil & Perryman, 2024; Lo, 2023; Tzanova, 2023). The demand for an AI literacy program for LIS professionals aligns with the rapidly changing technological environment in which libraries function and the increasing integration of artificial intelligence into various library services.

This study also found that there is a significant disparity between the perceived importance of AI competencies and the effectiveness of existing training programs, as was also highlighted in the study (Lo, 2024; Mallikarjuna, 2024). This is a major challenge, and demands an urgent need for targeted, tailormade, specialized practical AI literacy training for LIS professionals. The moderate satisfaction level with the current AI training programs suggests that there is tremendous room for improvement, both content-wise and delivery method, with more focus on practical and workshop-related training initiatives. In this study, ethical considerations emerged as one of the most crucial aspects of AI literacy (Diyaolu et al., 2024; Lo, 2024), and LIS professionals showed consensus regarding the concerns about privacy implications and the need to have clear ethical standards and guidelines for AI implementation in libraries, as highlighted in the study (Adewojo & Amzat, 2024; Kavak, 2024; Rajkumar et al., 2024). The study accentuates the dual roles of librarians in terms of implementers of AI technologies and as the ethical custodians of the use and access of information.

## Recommendations and Conclusion

This study has both theoretical and practical implications, as this study is distinctive in nature, as it tries to evaluate and assess cognitive, normative, and behavioral competencies related to AI in the field of library science, and is a unique addition to the literature encompassing artificial intelligence, librarianship, and social sciences. There is a substantial gap between the perceived importance of AI skills and competencies level and the effectiveness of the existing training programs in the UAE. The practical implication of this study recommends specialized, tailored intervention of AI training programs for library professionals, which need to be more practically oriented to enhance their learning. The study also highlights the need for libraries to adopt ethical standards and guidelines for the implementation of AI technologies. AI literacy and AI education need to be prioritized in professional development plans at organizational levels and should be integrated into the LIS curriculum in collaboration with computer science departments to understand the concept behind algorithms and coding patterns involved in generative AI at universities offering LIS courses around the world. Addressing these gaps can enhance AI implementation across all types of libraries and enable the library professionals to use AI tools and technologies effectively and in a responsible manner, addressing all ethical concerns and issues (Deshen & Aharony, 2024; Kizhakkethil & Perryman, 2024; Tzanova, 2024).

Another aspect of training is to equip the librarians with identify and mitigate AI biases. The study also recommends that universities and library associations come up with clear ethical standards and guidelines for AI use in libraries, which can address privacy concerns and ethical implications. Practical knowledge and hands-on experience with AI tools and technologies should be integrated into both formal education (LIS) and professional development programs. This study recommends investment and providing resources to support AI literacy initiatives in LIS education programs, as proposed by Chatikobo & Pasipamire (2024). One limitation of this study is that it is based on self-assessment of the library professionals, and it provides data about the perceived competencies and knowledge. This study does not cover an in-depth analysis of various training methods, and as such, there is a limited assessment of training effectiveness.

This study recommends future studies in measuring the effectiveness of training related to AI in libraries, assessing AI literacy levels among LIS professionals and LIS students through a framework, and a study involving the identification and measurement of practical applications of AI in library services and operations. This study provides a comprehensive foundation for understanding the current state and future needs of AI literacy in libraries and the Library & Information Science (LIS) profession. The findings of this study can serve as a valuable guide for Library Associations, professional organizations, and educational institutions in developing strategies to enhance AI literacy among library professionals.

## Acknowledgement

I would like to express my sincere gratitude to Dr. Michelle Demeter, Editor of College & Research Libraries, for taking the time to thoroughly review my work, cross-check the details, and provide insightful and constructive feedback that greatly helped improve the quality and clarity of the manuscript.